\documentclass[oneside, 12pt]{article}
\usepackage[english]{babel}
\usepackage{amssymb}
\usepackage{amsmath}
\usepackage{stmaryrd}
\usepackage{txfonts}
\usepackage{epsfig}
\usepackage{relsize}
\usepackage[symbol]{footmisc}
\newcounter{daggerfootnote}

\begin{document}

\hyphenation{ }
\vskip 1cm

\date{}
\begin{center}
    {\bf \Large Reaction-diffusion kinetics in growing domains}
\end{center}
\begin{center}
C. Escudero$^{1,}\footnote{Corresponding author; E-mail: cel@icmat.es}$, S. B. Yuste$^{2}$, E. Abad$^{3}$, F. Le Vot$^{2}$
\\[0.2cm]
$^{1}$ Departamento de Matem\'aticas \\
Universidad Aut\'onoma de Madrid \\
and Instituto de Ciencias Matem\'aticas \\
Consejo Superior de Investigaciones Cient\'{\i}ficas,\\
E-28049 Madrid, Spain \\
$^{2}$ Departamento de F\'{\i}sica and Instituto de Computaci\'on Cient\'{\i}fica Avanzada (ICCAEX) \\
  Universidad de Extremadura, E-06071 Badajoz, Spain \\
$^{3}$ Departamento de F\'{\i}sica Aplicada and Instituto de Computaci\'on Cient\'{\i}fica Avanzada (ICCAEX) \\ Centro Universitario de M\'erida \\ Universidad de Extremadura, E-06800 M\'erida, Spain \\
\end{center}

\section*{Abstract}
\hspace{0.6cm} Reaction-diffusion models have been used over decades to study biological systems. In this context, evolution equations for probability distribution  functions and the associated stochastic differential equations have nowadays become indispensable tools. In population dynamics, say, such approaches are utilized to study many systems, e.g., colonies of microorganisms or ecological systems. While the majority of studies focus on the case of a static domain, the time-dependent case is also important, as it allows one to deal with situations where the domain growth takes place over time scales that are relevant for the computation of reaction rates and of the associated reactant distributions. Such situations are indeed frequently encountered in the field of developmental biology, notably in connection with pattern formation, embryo growth or morphogen gradient formation. In this chapter, we review some recent advances in the study of pure diffusion processes in growing domains. These results are subsequently taken as a starting point to study the kinetics of a simple reaction-diffusion process, i.e., the encounter-controlled annihilation reaction. The outcome of the present work is expected to pave the way for the study of more complex reaction-diffusion systems of possible relevance in various fields of research.

\subsection*{Keywords} Random dispersal, reaction-diffusion kinetics, growing domains.

\section{Introduction}
\hspace{0.6cm} Reaction-diffusion equations are ubiquitous in the field of mathematical modeling of biological systems \cite{Murray2003}.
While in many instances the use of such models is justified, one should bear in mind that they are based on a mean-field description which completely neglects the effect of internal fluctuations characteristic of any discrete system. It is, however, well known that fluctuations may bring about drastic changes in the dynamics, resulting in a breakdown of mean-field approximations. Such fluctuation effects have been studied in many works relying on stochastic approaches, see e.g.
Refs.~\cite{Allam2013,Ben-Avraham1995b,Ben-Avraham1998c,Ben-Avraham1998b,Ben-Avraham2005,Burschka1989,Doering1989,Doering1991,Durang2011,Durang2014,Fortin2014,
Krapivsky2015,Lee1994,Lin1990,Lindenberg1995,Mandache2000,Masser2001a,Matin2015,Peliti1986,Sancho2007,Spouge1988,Torney1983,Winkler2012}.

Reaction-diffusion systems have been predominantly studied in static spatial domains. However, the case of a growing domain is of special relevance for biological systems; in particular, the interest of growing domains becomes self-evident in the context of developmental biology, where a common feature of the studied systems is their time-dependent size. This partly explains why reaction-diffusion processes in growing domains are currently the object of increasing interest \cite{Averbukh2014,Baker2010,Crampin1999,Crampin2001,Fried2015,Kulesa1996,Madzvamuse2010,Murray2003,Simpson2015d,Simpson2015b,Yuste2016}.
Despite several recent works that have already shed some light on the problem \cite{Angstmann2017,Simpson2015d,Simpson2015b,Yuste2016}, there are still many open questions and we dare say that this field of research has a promising future.

Here, we present some recent advances in the field of reaction-diffusion processes in growing domains. In light of previous works, we first review
the phenomenology of diffusion in growing domains in Sec.~\ref{diff}. In doing so, we first analyze the problem at the level of individual paths, suitably described by the pertinent stochastic differential equations. Following this, we derive the relevant diffusion equations associated with the spatial distribution of Brownian paths.

The actual purpose of the above approach to the problem of diffusion in growing domains is to pave the way for the study of reaction-diffusion processes. In
Sec.~\ref{sec:rq}, we focus on the mean-field kinetics of a specific reactive process, namely, the annihilation reaction  $A + A \rightarrow \emptyset$. This short-hand notation describes a process in which each collision between two identical particles leads to (instantaneous) annihilation with non-zero probability. This choice is by no means fortuitous, since the numerous previous studies on the kinetics of the (diffusion-controlled) annihilation reaction in static domains will be used as a starting point to treat the case of a growing domain.

For the sake of simplicity, we shall restrict ourselves to the case of uniform growth, implying that each volume element of the spatial domain expands at the same rate. Further, we assume that this uniform growth in time is described by a power law with a characteristic exponent $\gamma$ (also termed ``growth index'' hereafter). As the value of the growth index is varied, three different regimes are observed, namely, the subcritical case ($0<\gamma < 1$), the critical case ($\gamma=1$),
and the supercritical case ($\gamma >1$). We anticipate that the qualitative behavior in the subcritical case is essentially determined by the intrinsic reaction-diffusion process, implying that the domain growth can be effectively regarded as a perturbation that is only relevant from a quantitative point of view. In contrast, for sufficiently long times, the domain growth becomes the dominant effect in the supercritical case; in this limit the reaction-diffusion kinetics becomes irrelevant on a macroscopic scale. Finally, in the critical case, the observed marginal behavior arises from a subtle interplay between reaction, diffusion and domain growth.

Next, in Sec.~\ref{sec:ipi}, we investigate the role of fluctuations in the kinetics of the annihilation reaction. Fluctuations are known to be highly relevant in the static case, and the question is whether this also applies in the case of a growing domain. We find a certain agreement with the mean-field prediction, in the sense that the exact solution also predicts a transition between a regime where the concentration decreases due to the chemical reactions and another regime where the decay is determined by the dilution effect associated with the domain growth. However, the critical value of the growth index separating these two regimes is shifted to a lower value $\gamma_c=1/2$.

Finally, in Sec.~\ref{conclusions} we summarize our main findings and we comprehensively discuss the advantages and drawbacks of PDE approaches as a tool to study reaction-diffusion systems in growing domains.

\section{Diffusion on a uniformly growing domain}
\label{diff}

\hspace{.6cm} Consider the discrete-time evolution of a random walker on a one-dimensional domain which grows at a certain uniform rate. The present discussion is based on our previous study on the subject, see Ref.~\cite{Yuste2016} for more details.

\subsection{Langevin equation}

\hspace{0.6cm} Denoting by $y(t)$ the walker's position at time $t$, the stochastic evolution equation describing its motion reads as follows:
\begin{equation}\label{disc}
y(t + \Delta t) = \left( \frac{t + \Delta t}{t} \right)^\gamma y(t) + \sigma \, \Delta y_t \, \Delta t^{1/2},
\end{equation}
where $t \in \Delta t \times \mathbb{N}$, $\Delta t, \sigma \in \mathbb{R}^+$,
and the $\Delta y_t$'s are independent, identically distributed Gaussian random variables, i.e., $\mathcal{N}(0,1) \backsim \Delta y_t$. The above equation corresponds to the specific case where the time dependence of the uniform domain growth is prescribed by a power law with a characteristic exponent $\gamma$.
The importance of a power law domain growth was highlighted in~\cite{Yuste2016}. From a physical perspective, the mixing properties of the system suffer a drastic change in the vicinity of the critical value $\gamma=1/2$; in this regime, a competition between two opposite tendencies takes place, namely, diffusive mixing (which makes the typical separation between neighboring Brownian walkers decrease as $t^{1/2}$), and the deterministic drift induced by the domain growth, which makes the interparticle distance increase at a comparable rate ($\propto t^\gamma$). On the biological side, applications where the domain growth is described by a power law are indeed very frequent, such as the case of biological systems display a linear growth rate~\cite{Yates2014}.
The importance of this critical value becomes apparent already in this section.

In the limit $\Delta t \ll t$, Eq.~\eqref{disc} becomes
\begin{equation}\label{disc2}
y(t + \Delta t) \approx \left( 1 + \gamma \frac{\Delta t}{t} \right) y(t) + \sigma \, \Delta y_t \, \Delta t^{1/2},
\end{equation}
In particular, Eq.~\eqref{disc2} holds if we take $t \ge t_0$, where $t_0 = N \Delta t$ with $N \gg 1$.
Taking the continuum time limit in Eq.~\eqref{disc2} yields the stochastic differential equation
\begin{equation}\label{sde}
dy_t = \frac{\gamma}{t}y_t \, dt + \sigma \, dB_t.
\end{equation}
In what follows, we focus on the case $y(t_0)$ =0, i.e., the random walker is initially located at the origin. Here, ``initially'' means at time $t_0$.  The solution of the linear equation~\eqref{sde} is
\begin{equation}
y_t = \sigma t^\gamma \int_{t_0}^t s^{-\gamma} dB_s.
\end{equation}
As a consequence, $y_t$ represents a zero-mean Gaussian process, i.e.,
\begin{equation}
\mathbb{E}(y_t)=0,
\end{equation}
where we have used the martingale property of the It\^o integral. The covariance can be expressed as follows:
\begin{equation}
\label{coveq}
\mathbb{E}(y_t y_s)=
\begin{cases}
\sigma^2 t^\gamma s^\gamma \dfrac{(t \wedge s)^{1-2 \gamma} -t_0^{1-2\gamma}}{1- 2 \gamma} \quad \text{if} \quad  \gamma \neq \frac{1}{2} \\ \\
\sigma^2 t^{1/2} s^{1/2} \ln\left(\dfrac{t \wedge s}{t_0}\right) \quad \text{if} \quad \gamma = \frac{1}{2}
\end{cases}
,
\end{equation}
where the property of independent increments of the Brownian motion and the It\^o isometry have been used, as well as the notation $t \wedge s := \min\{t,s\}$.
Moreover, Eq.~\eqref{coveq} allows one to compute the variance, whence some intuition on the long-time behavior of the underlying dispersal process can be gained:
\begin{equation}\label{2moment}
\mathbb{E}(y_t^2) \approx
\begin{cases}
\dfrac{\sigma^2}{1- 2 \gamma} t \quad \text{if} \quad  \gamma < \frac{1}{2} \\ \\
\sigma^2 t \ln(t) \quad \text{if} \quad \gamma = \frac{1}{2} \\ \\
\dfrac{\sigma^2 t_0^{1-2\gamma}}{2 \gamma -1} t^{2 \gamma} \quad \text{if} \quad  \gamma > \frac{1}{2}
\end{cases}
.
\end{equation}
Thus, at the level of the mean squared displacement, a crossover between linear behavior and $\gamma$-dependent, superlinear behavior takes place as soon as the growth index exceeds the critical value $\gamma= 1/2$.

\subsection{Fokker-Planck equation}

\hspace{0.6cm} Our next goal will be to give an alternative description of the diffusion process studied in the previous subsection in terms of a partial differential equation. A major advantage of such a description is that it lends itself easily to the subsequent inclusion of reaction processes (see Secs.~\ref{sec:rq} and \ref{sec:ipi}).

For the time being, we continue to restrict ourselves to the case of a one-dimensional, uniformly growing  domain. However, given the importance of boundary effects in the theory of PDEs, we shall choose a slightly more general formulation by considering a finite interval of the form $[-L(t),L(t)]$.
Since the domain growth is uniform, one has $L(t)=a(t) L_0$, where $a(t) \ge 1$ will be hereafter termed the growth factor and $L_0\equiv L(0)$ stands for the initial semi-interval length (in order to recover the case of an infinite domain one lets $L_0$ go to infinity).
As shown in Ref.~\cite{Yuste2016}, the Fokker-Planck equation for the pdf $P(y,t)$ associated with the probability of finding the walker inside the infinitesimal interval $[y, y+dy]$ reads
\begin{equation}
\label{fpe}
\partial_t P(y,t)=-\partial_y \left[ \dfrac{\dot a(t)}{a(t)} y P(y,t) \right] + D \partial^2_y P(y,t).
\end{equation}
The first term on the right hand side can be split into two contributions describing the effects associated with the domain growth, namely, (1) a dilution term $-[\dot a(t)/a(t)] P(y,t)$ arising from a decay in particle concentration due to the increased system volume, and (2) a drift term $-[\dot a(t)/a(t)] y \partial_y  P(y,t)$ arising from the (deterministic) displacement of each volume element as it expands. As this stage, the change of variable
\begin{equation}\label{cc1}
x=\frac{L_0}{L(t)}y=\frac{y}{a(t)},
\end{equation}
turns out to be very convenient, since it maps the above transport problem into an equivalent problem on a fixed domain. The values of the so-called comoving coordinate $x$ are indeed restricted to the original interval $[-L_0,L_0]$ at all times.

In terms of the comoving coordinate, the pseudo-Fokker-Planck equation describing the time evolution of $P(x,t)\equiv P(y=a(t)x,t)$ is
\begin{equation}\label{d1}
    \partial_t P(x,t)=-\frac{\dot a(t)}{a(t)}P(x,t)+\frac{D}{a^2(t)} \partial^2_xP(x,t).
\end{equation}
In this representation, a dilution term arising from probability conservation still survives, but the drift term is now absent.

As already implicit in Eq.~\eqref{disc}, in what follows we shall focus on the case of a domain growth described by the following power-law:
\begin{equation}\label{d2}
    a(t)=\left(\frac{t}{t_0}\right)^{\gamma}, \qquad t \ge t_0,
\end{equation}
where $t_0 > 0$ is the initial time. Unless otherwise specified, the growth index $\gamma$ will take positive values (note, however, that the theory developed in the present work also applies in the case $\gamma<0$ of a shrinking domain). Then, Eq.~\eqref{d1} takes the form
\begin{equation}\label{d3}
    \partial_tP=-\frac{\gamma}{t}P+D\left(\frac{t}{t_0}\right)^{-2\gamma} \!\! \partial^2_xP.
\end{equation}
Consider now the initial value problem given by Eq.~\eqref{d3} and the set of equations
\begin{equation}\label{cf}
\begin{cases}
    P=P(x,t)\\
    P:\mathbb{R}\times[t=t_0,t=\infty)\rightarrow [0,\infty)\\
    P(x,t_0)=\delta(x)
\end{cases}
,
\end{equation}
where $\delta(x)$ denotes the Dirac delta function centered at the origin. This corresponds to the case of infinite system size ($L_0\to \infty$).
In order to solve Eq.~(\ref{d3}), we perform the following transformation:
\begin{equation}\label{cv}
    P(x,t)=\left(\frac{t}{t_0}\right)^{-\gamma}Q(x,t),
\end{equation}
which yields
\begin{equation}
\label{EqQxt}
    \partial_t Q(x,t)=\frac{D}{a^2(t)} \partial^2_x Q(x,t).
\end{equation}
In contrast with the auxiliary function $P(x,t)$, the quantity $Q(x,t)$ is a genuine pdf. As such, it is normalized and it has a well-defined physical meaning ($Q(x,t)\,dx$ is the probability to find the walker inside the interval $[x,x+dx]$ on the fixed domain). We draw the reader's attention to the time-dependence of the effective diffusion coefficient $D/a^2(t)$, arising from the fact that jumps become increasingly shorter in terms of the comoving coordinate $x$.

Following Ref. \cite{Yuste2016}, we now introduce the so-called Brownian conformal time $\tau=\int_{t_0}^t a(t')^{-2} \, dt'$. For the present case of a power-law expansion, we obtain:
\begin{eqnarray}\label{cv1}
 \tau=\begin{cases}
t_0^{2\gamma}\dfrac{t^{-2\gamma+1}-t_0^{-2\gamma+1}}{1-2\gamma} \quad \text{if} \quad  \gamma \neq \frac{1}{2} \\ \\
    t_0 \ln\left(\dfrac{t}{t_0}\right) \quad \text{if} \quad \gamma = \frac{1}{2}
\end{cases}
.
\end{eqnarray}
The conformal time allows one to absorb the growth factor into the time scale. Consequently, one has
\begin{equation}
\label{d4}
\partial_{\tau}Q(x,\tau)=D\partial^2_{x}Q(x,\tau),
\end{equation}
where
\begin{equation}\label{cf1}
\begin{cases}
    Q=Q(x,\tau) \\
    Q:\mathbb{R}\times[\tau=0,\tau=\infty)\rightarrow [0,\infty) \\
    Q(x,0)=\delta(x)
\end{cases}
,
\end{equation}
if $\gamma \le 1/2$ and
\begin{equation}
\begin{cases}
    Q=Q(x,\tau) \\
    Q:\mathbb{R} \times \left[\tau=0,\tau=\frac{t_0}{2\gamma-1} \right)\rightarrow [0,\infty) \\
    Q(x,0)=\delta(x)
\end{cases}
,
\end{equation}
if $\gamma >1/2$.
The well-known solution of Eq.~\eqref{d4} is
\begin{equation}\label{sd1}
    Q(x,\tau)=\frac{e^{-\frac{x^2}{4D\tau}}}{(4\pi D \tau)^{1/2}},
\end{equation}
whence the solution of Eq.~\eqref{d3} follows:
\begin{equation} \label{pexp}
P(x,t)=\\ \begin{cases}
\! \dfrac{1}{(4\pi D)^{1/2}} \! \left(\dfrac{t}{t_0}\right)^{-\gamma} \! \left[ t_0^{2\gamma}\dfrac{t^{-2\gamma+1}-t_0^{-2\gamma+1}}{1-2\gamma}\right]^{-1/2} \! \exp \! \left\{\!-\dfrac{x^2}{4D}\left[ t_0^{2\gamma}\dfrac{t^{-2\gamma+1}-t_0^{-2\gamma+1}}{1-2\gamma}\right]^{-1} \! \right\} \\ \\
\quad \text{if} \quad  \gamma \neq \dfrac{1}{2}, \\ \\
           \! \dfrac{1}{(4\pi D)^{1/2}} \! \left(\dfrac{t}{t_0}\right)^{-\gamma} \! \left[ t_0 \ln\left(\dfrac{t}{t_0}\right)\right]^{-1/2} \! \exp \! \left\{ \! -\dfrac{x^2}{4D}\left[ t_0 \ln\left(\dfrac{t}{t_0}\right)\right]^{-1} \! \right\} \\ \\ \quad \text{if} \quad \gamma = \dfrac{1}{2}.
\end{cases}
\end{equation}

We can use this explicit expression to derive the first-order moment and the second-order moment of $Q(x,t)$ [respectively denoted by $M_1$ and $M_2$]. We obtain
    \begin{equation}\label{m1}
M_1=\int_{\mathbb{R}} x \, Q(x,t) \,dx= \int_{\mathbb{R}} x \, P(x,t) \left(\frac{t}{t_0}\right)^{\gamma} \!\! dx=0,
\end{equation}
and
    \begin{equation}\label{m2}
M_2=\int_{\mathbb{R}} x^2 \, Q(x,t) \, dx=\int_{\mathbb{R}} x^2 P(x,t) \left(\frac{t}{t_0}\right)^{\gamma} \!\! dx=2D\tau.
\end{equation}
The result for $M_1$ follows directly from the system's symmetry; the second-order moment $M_2$ can be expressed in terms of the original time scale as follows:
\begin{equation}\label{cv11}
M_2=\begin{cases}
2 D t_0^{2\gamma}\dfrac{t^{-2\gamma+1}-t_0^{-2\gamma+1}}{1-2\gamma} \quad \text{if} \quad \gamma \neq \dfrac{1}{2} \\ \\
2 D t_0 \ln\left(\dfrac{t}{t_0}\right) \quad \text{if} \quad \gamma = \dfrac{1}{2}
\end{cases}
.
\end{equation}
In the limit $t\rightarrow \infty$, the above expressions become more transparent:
\begin{equation}\label{eqcm2}
M_2 \approx \begin{cases}
2D \dfrac{t_0^{2\gamma}}{1-2\gamma}t^{1-2\gamma}\propto t^{1-2\gamma} \quad \text{if} \quad \gamma < \dfrac{1}{2} \\ \\
2D t_0 \ln\left(t\right)\propto \ln(t) \quad \text{if} \quad \gamma = \dfrac{1}{2} \\ \\
2D \dfrac{t_0}{2\gamma-1}= \text{constant} \quad \text{if} \quad \gamma > \dfrac{1}{2}
\end{cases}
.
\end{equation}
In words, for $\gamma<\frac{1}{2}$ the intrinsic diffusion process dominates over the transport process due to the domain growth. On the other hand, when $\gamma>\frac{1}{2}$, the diffusive transport becomes negligible with respect to the particle spreading induced by the domain growth; as a result of this, the walker's jumps are strongly shortened in terms of the comoving coordinate, to the extent that the second-order moment tends to a constant in the long-time limit. Finally, for $\gamma = \frac{1}{2}$, the interplay between the domain growth and the intrinsic diffusive transport results in the latter being marginally faster. In this case, the slow spreading of the sojourn pdf $Q(x,t)$ is described by a logarithmic time growth of its second-order moment.

In terms of the physical coordinate, the long-time asymptotics of the second-order moment $\mathcal{M}_2=\int_{\mathbb{R}} y^2 \, P(y,t) \,dy$  follows straightforwardly from the general relation $\mathcal{M}_n=\int_{\mathbb{R}} y^n \, P(y,t) \,dy= a(t)^n M_n$ for any positive integer $n$. One has
\begin{equation}\label{eqcm3}
\mathcal{M}_2 \approx \begin{cases}
\dfrac{2D}{1-2\gamma} t \quad \text{if} \quad \gamma < \dfrac{1}{2} \\ \\
2D t \ln\left(t\right) \quad \text{if} \quad \gamma = \dfrac{1}{2} \\ \\
\dfrac{2D t_0^{1-2\gamma}}{2\gamma-1} t^{2\gamma} \quad \text{if} \quad \gamma > \dfrac{1}{2}
\end{cases}
.
\end{equation}
Let us briefly discuss the physics underlying the above formula. Below the threshold value $\gamma=1/2$, the diffusive transport dominates over the domain growth and determines the qualitative time dependence; there is only a weak signature of the growth process at the level of the effective diffusion coefficient, which becomes a monotonically increasing function of $\gamma$. Above  $\gamma=1/2$, the deterministic domain growth accelerates the particle spreading with respect to the case of a static domain. Finally, in the marginal case $\gamma=1/2$ the effect of the domain growth consists in introducing a logarithmic correction to the linear behavior in $t$ of the mean squared displacement.

Note that, with the identification $D \leftrightarrow \sigma^2/2$, Eq.~\eqref{eqcm3} and  Eq.~\eqref{2moment} are in full agreement with one another. A simple calculation show that this agreement goes beyond the first- and second-order moments and extends to the full set of moments.

Finally, assume that we wish to extend the above solution to the case of an arbitrary initial condition in a $d$-dimensional hypercubic domain $[-L(t),L(t)]^d$. In the limit $L_0\to \infty$, the solution to the counterpart of the initial value problem~\eqref{d3}-\eqref{cf} with an arbitrary initial condition $P(\textbf{x},t_0)=g(\textbf{x})$ is given by the convolution of $g(\textbf{x})$ with the corresponding $d$-dimensional propagator, i.e.,
\begin{subequations}
\label{sd2g}
\begin{align}
P(\textbf{x},t)=&
 \dfrac{t^{-d \gamma}}{(4\pi D)^{d/2}} \left[\dfrac{t^{-2\gamma+1}-t_0^{-2\gamma+1}}{1-2\gamma}\right]^{-d/2} \nonumber \\ &\times
\mathlarger{\mathlarger{\int}}_{\mathbb{R}^d}
\exp  \left\{  -\dfrac{(\textbf{x}-\textbf{x}')^2}{4D}\left[ t_0^{2\gamma}\dfrac{t^{-2\gamma+1}-t_0^{-2\gamma+1}}{1-2\gamma}\right]^{-1}  \right\}
g(\textbf{x}') \, d\textbf{x}'
\end{align}
 if $\gamma \neq 1/2$ and
\begin{equation}
P(\textbf{x},t)=
             \dfrac{1}{(4\pi D)^{d/2}} \left[ t \ln\left(\dfrac{t}{t_0}\right) \right]^{-d/2}
          \! \mathlarger{\mathlarger{\int}}_{\mathbb{R}^d} \exp  \left\{  -\dfrac{(\textbf{x}-\textbf{x}')^2}{4D}\left[ t_0 \ln\left(\dfrac{t}{t_0}\right)\right]^{-1}  \right\}
           g(\textbf{x}') \, d\textbf{x}'
\end{equation}
\end{subequations}
 if $\gamma = 1/2$.
 Let us consider the special case $g(x)=\frac{1}{2}e^{-|x|}$ in one spatial dimension. We obtain
\begin{subequations}
\label{sd2g2}
\begin{align}
P(x,t)=&
\frac{1}{4} \left(\frac{t}{t_0}\right)^{-\gamma}
\! \exp \! \left[ \! -\frac{x^2}{4D}\left( t_0^{2\gamma}\frac{t^{-2\gamma+1}-t_0^{-2\gamma+1}}{1-2\gamma}\right)^{-1} \! \right] \nonumber \\
&\times
\left\{ \exp \left[ \frac{\left(D t_0^{2\gamma}\frac{t^{-2\gamma+1}-t_0^{-2\gamma+1}}{1-2\gamma} -\frac{x}{2} \right)^2}
{D t_0^{2\gamma}\frac{t^{-2\gamma+1}-t_0^{-2\gamma+1}}{1-2\gamma}} \right]
\mathrm{erfc} \left(  \frac{D t_0^{2\gamma}\frac{t^{-2\gamma+1}-t_0^{-2\gamma+1}}{1-2\gamma} -\frac{x}{2}}
{\sqrt{ D t_0^{2\gamma}\frac{t^{-2\gamma+1}-t_0^{-2\gamma+1}}{1-2\gamma}}} \right) \right.
\nonumber
\\
& \;\;\quad + \left.
\exp \left[ \frac{\left(D t_0^{2\gamma}\frac{t^{-2\gamma+1}-t_0^{-2\gamma+1}}{1-2\gamma} +\frac{x}{2} \right)^2}
{D t_0^{2\gamma}\frac{t^{-2\gamma+1}-t_0^{-2\gamma+1}}{1-2\gamma}} \right]
\mathrm{erfc} \left(  \frac{D t_0^{2\gamma}\frac{t^{-2\gamma+1}-t_0^{-2\gamma+1}}{1-2\gamma} +\frac{x}{2}}
{\sqrt{ D t_0^{2\gamma}\frac{t^{-2\gamma+1}-t_0^{-2\gamma+1}}{1-2\gamma}}} \right)
\right\}
\end{align}
 if $\gamma \neq 1/2$ and
\begin{align}
P(x,t)=&
           \frac{1}{4} \left(\frac{t}{t_0}\right)^{-\gamma}
           \exp\! \left[  -\frac{x^2}{4D}\left( t_0 \ln\left(\frac{t}{t_0}\right)\right)^{-1} \! \right] \nonumber \\
           &\times
           \left\{ \exp \left[ \frac{\left(D t_0 \ln\left(\frac{t}{t_0}\right) -\frac{x}{2} \right)^2}
           {D t_0 \ln\left(\frac{t}{t_0}\right)} \right]
           \mathrm{erfc} \left( \frac{D t_0 \ln\left(\frac{t}{t_0}\right) -\frac{x}{2}}
           {\sqrt{ D t_0 \ln\left(\frac{t}{t_0}\right)}} \right)  \right.
           \nonumber
           \\
           & \;\;\quad +\left.
           \exp \left[ \frac{\left(D t_0 \ln\left(\frac{t}{t_0}\right) +\frac{x}{2} \right)^2}
           {D t_0 \ln\left(\frac{t}{t_0}\right)} \right]
           \mathrm{erfc} \left(  \frac{D t_0 \ln\left(\frac{t}{t_0}\right) +\frac{x}{2}}
           {\sqrt{ D t_0 \ln\left(\frac{t}{t_0}\right)}} \right)
           \right\}
\end{align}
 if $\gamma =1/2$.
\end{subequations}

\begin{figure}[t]
\begin{center}
        \includegraphics[width=0.7\textwidth,angle=0]{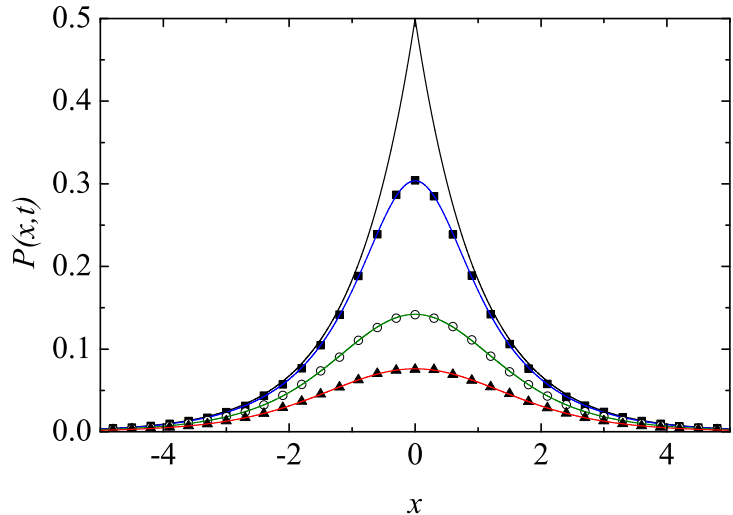}
\end{center}
\caption{\label{fig1} Time evolution of $P(x,t)$ for an exponentially decaying initial condition. We have set $\gamma=3/4$ and $t_0=10$. Solid curves represent analytic results; each set of dots represents simulation results obtained from $10^6$ runs.  Each curve corresponds, from top to bottom to the values $t={t_0, 12.5, 25, 50}$. The intrinsic random walk parameters are $\Sigma=\sqrt{10^{-3}}$ and $\langle t \rangle=0.01$, resulting in the diffusion coefficient $D=0.05$ (see main text).}
\end{figure}

In Fig. \ref{fig1} we compare the analytic result \eqref{sd2g2} with the outcome of numerical simulations. The intrinsic motion of the particles has been implemented by means of an uncoupled Continuous-Time-Random-Walk (CTRW) model with an exponentially decaying waiting time pdf $\psi(t)\propto \exp{(-t/\langle t\rangle)}$ and a Gaussian jump length pdf $\lambda(\Delta y)\propto \exp{(-|\Delta y|^2/2\Sigma^2)}$. On a static domain, this model is known to yield normal diffusion with diffusion coefficient $D= \Sigma^2/ (2\langle t \rangle)$. One can see that the agreement between theory and simulations is excellent.

\section{Encounter-controlled annihilation: mean-field theory}
\label{sec:rq}

\hspace{.6cm} Having studied the statistical properties of pure diffusion processes, we now turn our attention to one of the simplest reactive processes driven by binary collisions, i.e., the encounter-controlled annihilation reaction
                 $$A + A \stackrel{\frac{\alpha}{2}}{\longrightarrow} \emptyset.$$
In words, a pair of homologous particles $A$ interact with one another upon encounter and disappear instantaneously at a Poisson rate $\alpha/2$ as a result of the interaction. That is, we are considering the spatially extended version of a continuous-time Markov chain (details on the theoretical framework for continuous-time Markov chains can be found in the references \cite{Durrett} and \cite{Ross}). In this context, we remind the reader that in the limiting
case of a well-stirred chemical system (an effectively zero-dimensional system), diffusive mixing is not at play, as a result of which the annihilation events
are independent and the waiting time between two consecutive reactions is an exponentially distributed random variable whose characteristic parameter $\alpha/2$ determines the rate constant of the chemical reaction \cite{McI,McII}.

The above reaction has been comprehensively studied in the diffusion-controlled case. Denoting once again the diffusion coefficient of the particles by $D$, the mean-field description of the above process in the one-dimensional, diffusion-limited case is given by the set of equations
\begin{eqnarray}
\label{eqnes}
\begin{cases}
    \partial_tc = D \, \partial_y^2 c-\alpha c^2\\
    c(y,0) = C > 0
\end{cases}
,
\end{eqnarray}
where $c=c(y,t)$ is the (local) particle density. For simplicity, we shall restrict ourselves to the case where the initial density profile is constant, implying that local densities become identical with global densities given by spatial averages.

Clearly, Eqs.~\eqref{eqnes} correspond to a description where fluctuation effects are totally absent. Our goal in the present section will be to extend such a description to the case of a growing domain and to characterize the long-time behavior of the mean-field prediction. In Sec.~\ref{sec:ipi}, we shall compare this prediction with an exact solution based on an interval method. The latter takes full account of stochastic fluctuations arising from the chemical reactions. As a result of this, the long-time behavior of the exact solution will be seen to significantly differ from the mean-field prediction.

If we consider the initial value problem (\ref{eqnes}) on the real line $\mathbb{R}$ or on a finite ring, translational invariance is preserved at all times, and the spatial dependence of the particle density vanishes. Consequently, the Laplacian term on the right hand side of the first equation in (\ref{eqnes}) vanishes and one is left with the following problem:
\begin{eqnarray}\label{eqes}
\begin{cases}
    \dfrac{d c}{dt} = -\alpha c^2\\
    c(0) = C > 0
\end{cases}
.
\end{eqnarray}
The solution is
\begin{equation}
    c = \frac{C}{1+ \alpha \, C \, t}.
\end{equation}
For long times we see that
\begin{equation}
    c \approx \frac{1}{\alpha \, t} \propto \frac{1}{t},
\end{equation}
i.~e., the asymptotic behavior of the density of reagents displays a universal algebraic decay in time.

Next, let us consider the situation where the one-dimensional domain grows uniformly in the course of time. To obtain the relevant mean-field description, one performs the replacement $P(y,t)\to c(y,t)$ in Eq. \eqref{fpe}, and subsequently augments the right hand side of the resulting equation with the nonlinear reaction term $-\alpha \, c(y,t)^2$. This yields
\begin{equation}
\label{fpewithreact}
\partial_t c(y,t)=- \frac{\dot a(t)}{a(t)} c(y,t) - \frac{\dot a(t)}{a(t)} y \partial_y c(y,t) + D \partial^2_y c(y,t)-\alpha \, c(y,t)^2,
\end{equation}
that is, the time evolution of the concentration (term on the left hand side of this equation) is dictated by the diffusion of the reagents in the growing
domain (first three terms on the right hand side, see comments in the previous section) acting together with the annihilation reaction (fourth term on the right hand side, see~\cite{Ben-Avraham2005}). The proportionality of the latter to the square of the concentration is a signature of the mean-field assumption, implying that for moderate expansion rates the system is locally well mixed, and every particle in a given volume is able to interact with any other particle (the reaction rate is therefore proportional to the number of reactive pairs, which in turn grows approximately as the square of the number of particles). As the expansion rate becomes larger, the last two terms on the right hand side quickly become irrelevant; in this regime, the kinetics is dominated by the first two terms, which describe the drift-dilution effects induced by the domain growth.

For a spatially uniform initial condition $c(y,t_0)\equiv c_0$ and a translationally invariant, uniformly growing domain, one has $c(y,t)=c(t)$; consequently, the relevant initial value problem is
\begin{equation}
\label{eq1}
\begin{cases}
    \dfrac{dc}{dt} = -\dfrac{\dot a}{a}c -\alpha c^2 \\
    c(t_0)=c_0
\end{cases}
.
\end{equation}
We now once again focus on the special case of the power-law growth factor given by Eq.~\eqref{d2}. In this case, one has
\begin{equation}
\label{eq1p}
\begin{cases}
    \dfrac{dc}{dt} = -\dfrac{\gamma}{t}c -\alpha c^2 \\
    c(t_0)=c_0
\end{cases}
.
\end{equation}
Note that, contrary to Eq.~\eqref{eqes}, the above equation is not invariant with respect to a temporal shift. Consequently, we cannot choose arbitrarily the origin of time $t=0$, since this instant is singled out by its physical meaning. In fact, $t=0$ describes the singular limit in which the spatial domain shrinks to a point.

In order to prevent such a singularity, we restrict ourselves to the solution in the time domain $t >t_0$. Eq.~\eqref{eq1p} is still of Bernoulli type; consequently, it can be explicitly solved as well. Depending on the growth index $\gamma$, one of the two following cases applies:
\begin{subequations}
\begin{enumerate}
    \item If $\gamma \neq 1$, then
\begin{equation}\label{eqs11}
    c(t)= \frac{c_0 (1-\gamma t_0^{\gamma})}{c_0 \alpha \, t \, t_0^{\gamma} + t^{\gamma}(1-\gamma -c_0 \alpha t_0)}.
\end{equation}
    \item If $\gamma = 1$, then
\begin{equation}\label{eqs22}
     c(t)=\frac{c_0 t_0}{t + c_0 \alpha \, t \, t_0 \ln(t/t_0)}.
\end{equation}
\end{enumerate}
\end{subequations}
Taking the limit $t \rightarrow \infty$ in the explicit solutions (\ref{eqs11}) and (\ref{eqs22}), one can distinguish various subcases with different long-time behaviors:
\begin{enumerate}
    \item $\gamma \neq 1$
\begin{description}
    \item[a.] For $\gamma < 1$ and $\alpha\neq 0$ one finds
    \begin{equation}\label{eqc1a}
    c(t)\approx \frac{1-\gamma}{\alpha t}\propto t^{-1}.
    \end{equation}
    \item[b.] For $\gamma < 1$ and $\alpha= 0$, Eq. (\ref{eq1}) yields
\begin{equation}
    \partial_t c=-\frac{\gamma}{t}c,
\end{equation}
whose explicit solution reads
    \begin{equation}\label{eq3}
    c(t)=c_0\left(\frac{t}{t_0}\right)^{-\gamma}.
    \end{equation}
Consequently,
    \begin{equation}\label{eqc2}
    c(t)\propto t^{-\gamma}
    \end{equation}
(in fact, the above result holds for arbitrary $\gamma >0$ as long as $\alpha=0$).
    \item[c.] For $\gamma > 1$ one finds
\begin{equation}\label{eqc3}
    c(t)\approx \frac{c_0 (\gamma -1) t_0^{\gamma}}{\gamma -1 +c_0 \alpha t_0} t^{-\gamma}\propto t^{-\gamma}.
\end{equation}
\end{description}
    \item If $\gamma=1$ then
\begin{equation}\label{eqc4}
     c(t)\approx \left[\alpha \, t \, \ln\left(t\right)\right]^{-1}\propto \frac{1}{t \, \ln(t)}.
\end{equation}
\end{enumerate}
These results allow us to distinguish the following regimes:
\begin{itemize}
\item If $\alpha =0$ and $\gamma >0$ then $c \propto t^{-\gamma}$. That is, in the absence of annihilation the density decreases by dilution.
\item If $\gamma =0$ and $\alpha >0$ then $c \propto t^{-1}$. That is, in the absence of dilution the density decreases by annihilation.
\item If $\alpha >0$ and $\gamma >0$ we distinguish the following subcases:
\begin{itemize}
    \item If $\gamma<1$ then $c \propto t^{-1}$. That is, if the growth rate is small enough ($\gamma < 1$) the decrease in density is
    driven by annihilation.
    \item If $\gamma>1$ then $c \propto t^{-\gamma}$. That is, if the growth rate is large enough ($\gamma > 1$) the decrease in density is
    driven by dilution.
    \item If $\gamma=1$ then $c \propto [t \, \ln(t)]^{-1}$. This marginal case is interesting, since the decrease in density is slightly faster
    than in those cases for which either $\alpha=0$ (no annihilation) or $\gamma=0$ (no dilution). Both the domain growth and the annihilation reaction contribute significantly to the density decay, and there is no clear prevalence of one effect over the other.
\end{itemize}
\end{itemize}

\section{Exact solution for encounter-controlled annihilation}
\label{sec:ipi}

\hspace{0.6 cm} As it is well known, for diffusion-limited annihilation on a static domain, $d_c=2$ is the critical dimension separating mean-field behavior for $d>d_c$ from fluctuation-dominated decay for $d<d_c$. While fluctuation effects are only marginal in $d=2$, such effects are much stronger in $d<2$. In particular, fluctuation effects are very relevant in one dimension. For an infinite reaction rate, this case is amenable to exact solution via the even/odd interval method \cite{Masser2001a}.  The breakdown of the mean-field approach in the case of a static domain suggests that similar effects take place on a growing domain, at least for sufficiently low growth rates. Fortunately, the exact solution available for the standard case of a static domain can be straightforwardly extended to the case of a uniformly growing domain.

In what follows, we shall construct the exact solution by using the interval method originally employed in Ref.~\cite{Masser2001a}. Let us denote by $G(y,t)$ the probability that an arbitrarily chosen interval of length $y$ contains an even number of particles at time $t$. This quantity can be shown to obey a pure diffusion equation in the static case. On a growing domain, the only difference is that this equation must be augmented with an additional drift term. For particles with diffusion coefficient $D$, we have
\begin{equation}
\label{1.4}
    \frac{\partial}{\partial t} G(y,t) + \frac{\dot a}{a} y \partial_y G(y,t) =2D\partial_y^2 G(y,t),
\end{equation}
where $y \in [-a(t) L_0, a(t) L_0]$ (eventually we will take the limit $L_0 \to \infty$). Note that the above equation differs from Eq.~(\ref{fpe}) in that the latter contains not only a drift term, but also a dilution term. The reason is that Eq.~\eqref{fpe} describes the evolution of a probability \emph{per unit volume}, whereas Eq.~\eqref{1.4} refers to a probability. Therefore, a dilution term due to volume growth appears in the former case, but not in the latter.

Let us now introduce comoving coordinates by performing the replacement
\begin{equation}\label{1.3}
\left\{
\begin{array}{l}
y=a(t) \, x \\
dy=a(t) \, dx
\end{array}
\right.
\end{equation}
in Eq.~(\ref{1.4}). This yields
\begin{equation}\label{1.5}
    \frac{\partial}{\partial t} G(x,t)= \frac{2D}{a^2(t)}\frac{\partial^2}{\partial x^2} G(x,t).
\end{equation}
In the limit $L_0\to \infty$, this equation is to be solved subject to the following conditions:
\begin{equation}\label{1.6}
\left\{
\begin{array}{cr}
G(0,t)=1, \\
G(\infty,t)=\frac{1}{2}, & \hspace{1.0cm} 0<x<\infty, t\geq t_0 > 0,  \\
G(x,t_0)=G_0(x), & \hspace{1.0cm} 0 \le G_0(x) \le 1.
\end{array}
\right.
\end{equation}
We take as initial condition $G_0(x)=\frac{1}{2}+\frac{1}{2}e^{-2 c_0 x}$, which corresponds to a random homogeneous particle distribution with
mean concentration $c_0$ (this quantity is implicitly assumed to be small enough so as to ensure the validity of the above approach in the diffusion-controlled regime). The problem posed by Eqs.~\eqref{1.5}-\eqref{1.6} can be cast into an equivalent problem with a time-independent diffusion coefficient. To do so, one can again use the definition of the Brownian conformal time $\tau$ to absorb the effect of the domain growth into the time scale
[cf. Eqs.~\eqref{EqQxt}-\eqref{d4}]. The solution reads
\begin{eqnarray}
G(x,\tau)&=&
\frac{1}{2}+\frac{1}{2}\text{erfc}\left(\frac{x}{\sqrt{8D \tau}}\right)-\frac{1}{4}e^{8D c_0^2 \tau} \\ \nonumber &&
\times \left\{ e^{2 c_0 x} \left[1-\text{erf}\left(\frac{x+8D c_0 \tau}{\sqrt{8D \tau}}\right) \right]-e^{-2 c_0 x}
\left[1+\text{erf}\left(\frac{x-8 D c_0 \tau}{\sqrt{8D \tau}}\right) \right]\right\}.
\end{eqnarray}

The particle concentration (particle number per unit length of the physical coordinate $y$) is then given by
\begin{equation}\label{1.11}
    c(t)=-\frac{1}{a(t)}\left.\frac{\partial}{\partial x}G[x,\tau(t)]\right|_{x=0}.
\end{equation}
Note the additional prefactor $a(t)^{-1}$ on the right hand side with respect to the static case described in Ref.~\cite{Masser2001a}.

Let us now once again consider the special case of a domain growth driven by the power-law $a(t)=(t/t_0)^\gamma$, with $\tau(t)$ given by Eq.~\eqref{cv1}.
Eq.~\eqref{1.11} then yields
\begin{eqnarray}
c(t) &=& \\ && \nonumber
\begin{cases}
c_0 \left( \frac{t}{t_0} \right)^{-\gamma} \exp \left\{ \frac{8D c_0^2 \left[ t \left( \frac{t}{t_0} \right)^{-2\gamma}
-t_0 \right]}{1-2\gamma} \right\} \mathrm{erfc}
\left\{ c_0 \sqrt{ \frac{8D \left[ t \left( \frac{t}{t_0} \right)^{-2\gamma}
-t_0 \right]}{1-2\gamma}} \right\}
\quad \text{if} \quad \gamma \neq \dfrac{1}{2} \\
c_0 \left( \frac{t}{t_0} \right)^{-\gamma} \exp \left[ 8D c_0^2  t_0 \ln \left( \dfrac{t}{t_0} \right) \right] \mathrm{erfc}
\left[ c_0 \sqrt{8D t_0 \ln \left( \dfrac{t}{t_0} \right)} \right]
\quad \text{if} \quad \gamma = \dfrac{1}{2} \\
\end{cases}
.
\end{eqnarray}

The above expressions are somewhat cumbersome, but they take a much simpler form in the long-time limit. According to the qualitative asymptotic time dependence, one can distinguish three different cases, namely:
\begin{itemize}
    \item $\gamma < \frac{1}{2}$
\begin{equation*}
    c(t) \approx \sqrt{\frac{1-2\gamma}{8 \pi D t}} \propto t^{-1/2},
\end{equation*}
    \item $\gamma = \frac{1}{2}$
\begin{equation}\label{1.14}
    c(t) \approx \frac{1}{\sqrt{8 \pi D t \ln (t/t_0)}} \propto [t \ln(t)]^{-1/2},
\end{equation}
    \item $\gamma > \frac{1}{2}$
    \begin{equation*}
    c(t) \approx c_0 \left( \frac{t}{t_0} \right)^{-\gamma} \exp \left( \frac{8 D c_0^2 t_0}{2 \gamma -1} \right)
    \mathrm{erfc}
    \left( c_0 \sqrt{\frac{8 D t_0}{2\gamma -1}} \right)
    \propto t^{-\gamma}.
\end{equation*}
\end{itemize}

\begin{figure}
\begin{center}
\includegraphics[width=0.7\textwidth,angle=0]{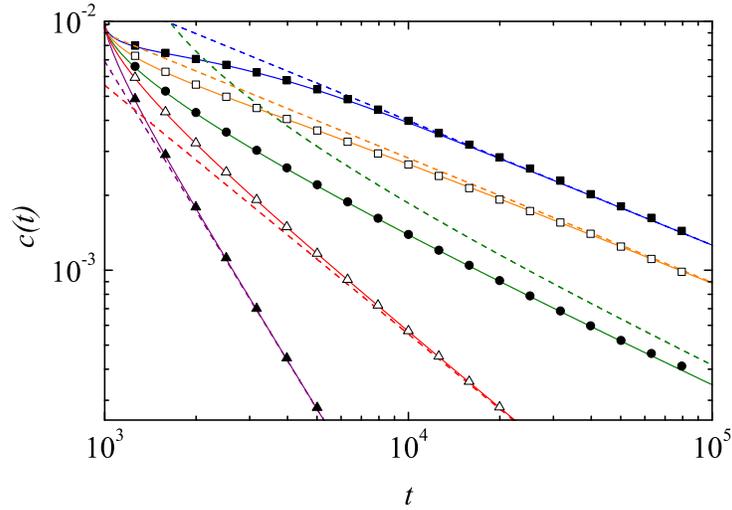}
\end{center}
\caption{\label{fig2} Time evolution of the mean concentration for different growth rates. Solid lines represent analytic results, whereas each set of dots correspond to simulation results obtained from $10^5$ runs. Dashed lines correspond to the long-time prediction based on Eqs.~\eqref{1.14}. We have chosen the initial values $c_0=0.01$ and $t_0=1000$. The diffusion coefficient resulting from the parameter choice of our CTRW model is $D=1/2$. The values of the growth index are, from top to bottom, $\gamma={-0.5,0,0.5,1,2}$. Note that the case $\gamma=-0.5$ corresponds to a shrinking domain. As already noted, our theory is also valid for this case.}
\end{figure}
In order to confirm the above analytic results, we have carried out CTRW simulations (cf. Sec.~2). Once again, we obtain excellent agreement between theory and simulations (see Fig.~\ref{fig2}). Both the ongoing reactions and the domain growth contribute to the density decay; at long times, the number of reactions decreases drastically for a sufficiently fast expansion ($\gamma>1/2$), and the dilution effect due to the domain growth prevails. In contrast, for $\gamma<1/2$ the decrease in the reaction rate induced by the domain growth is not sufficient to change the qualitative time dependence observed in the static case.

Finally, let us compare the above results with the mean-field prediction given in Sec.~\ref{sec:rq}. In both cases, we have three different regimes, namely: a subcritical regime in which the concentration decays due to particle annihilation, a supercritical regime in which the concentration decays due to dilution and a critical regime in which the density both dilution and reaction determine the density decay (the decay law contains a logarithmic term in the latter case). However, the crossover value of the growth index is shifted from the mean-field value $\gamma=1$  to $\gamma=1/2$. The reason is that the poor mixing of the reactants due to the constrained one-dimensional geometry lowers the reaction rate significantly and makes the dilution effect arising from the domain growth the dominant contribution to the density decay, even for moderate values of the growth index ($1/2<\gamma<1$). Only for $\gamma \ge 1$ does the decay law given by the exact solution become identical with the mean-field prediction. In contrast, for $\gamma<1$, the mean-field prediction fails to correctly describe the observed time dependence.

Finally, we note that one can also consider the evolution of the density decay. Denoting by $c_f(t)$ the density with respect to the fixed original domain, one has
\begin{equation}
c_f(t)= a(t)c(t) \propto \begin{cases}
t^{\gamma -1/2} \quad \text{if} \quad \gamma < 1/2 \\
[\ln(t)]^{-1/2} \quad \text{if} \quad \gamma = 1/2 \\
\text{constant}>0 \quad \text{if} \quad \gamma > 1/2
\end{cases}
.
\end{equation}
This quantity does not account for dilution arising from the domain growth, since the number of particles in a given interval is divided by the interval length at $t=0$. Therefore, if no reactions take place,  $c_f(t)$ remains constant. This is the case as soon as $\gamma>1/2$. In the long time regime, interparticle gaps arising from reactions at earlier times are strongly expanded by the ongoing domain growth; encounters between particles become rare, and the number of particles in the system with respect to the initial system length eventually stabilizes. In the opposite case $\gamma<1/2$, the decay of the total ``system mass'' is algebraic and progressively slows down as the growth index approaches from below the critical value $\gamma=1/2$, where the decay is only logarithmic.

\section{Conclusions and outlook}
\label{conclusions}

\hspace{0.6cm}
To begin with, in Sec.~\ref{diff} we have reviewed the phenomenology of diffusion processes on a growing, one-dimensional domain. Specifically, we have considered the case of a power-law growth factor $a(t)=(t/t_0)^\gamma$. A suitable choice of transformation allows one to map this problem onto a pure diffusion problem on a fixed domain. However, the interplay between the domain growth and the intrinsic diffusive transport may give rise to the onset of different regimes. The behavior is perhaps best summarized by recalling our results for the second-order moment taken with respect the physical coordinate $y$, i.e.,
\begin{equation}
\label{PhysMom}
\mathcal{M}_2 \approx \begin{cases}
2 D \frac{1}{1-2\gamma}t \propto t \quad \text{if} \quad \gamma < \frac{1}{2} \\
2 D t \ln\left(t\right)\propto t \ln(t) \quad \text{if} \quad \gamma = \frac{1}{2} \\
2 D \frac{t_0^{1-2\gamma}}{2\gamma-1} t^{2\gamma} \propto t^{2 \gamma} \quad \text{if} \quad \gamma > \frac{1}{2}
\end{cases}
.
\end{equation}
The above result can be obtained either from the Langevin equation describing the stochastic dynamics of individual diffusional paths or from a coarse-grained description relying on a pseudo-Fokker-Planck equation.
Eq.~\eqref{PhysMom} shows that there is a crossover value $\gamma=1/2$ separating the diffusion-controlled regime ($\gamma <1/2$) from the dilution-controlled regime ($\gamma >1/2$).

To some extent, the above results give a hint on the behavior of the exact solution for the encounter-controlled annihilation reaction $A+A \rightarrow \emptyset$ in one spatial dimension. To the best of our knowledge, this reaction-diffusion system is one of the few amenable to exact solution on a growing domain. As we have seen, a crossover between a diffusion-controlled regime and a dilution-controlled regime is also observed at $\gamma=1/2$. When $\gamma>1/2$, the domain growth is so fast that it prevents the system from becoming empty, and a number of unreacted particles survive forever.

As shown in Sec.~\ref{sec:rq}, the mean-field approach is able to capture the onset of different regimes, but its prediction for the time decay of the density is only correct in the limit of a fast domain growth $\gamma\ge 1$. Below this threshold value, the mean-field approximation breaks down, since the slowing-down of the kinetics due to fluctuations induced by the ongoing chemical reactions is enhanced by the dilution effect due to the domain growth.

In spite of the relevance of fluctuation effects in $d=1$, the fact that the mean-field approach is able to reproduce the correct qualitative behavior in the limiting case of a sufficiently fast growing domain is quite remarkable. This shows that a sufficiently fast domain growth may lower the critical dimension for the crossover to mean-field behavior, which is known to be $d=2$ in the case of a static domain \cite{Kamenev}.

In view of the above findings, it should be clear that the kinetics of diffusion-controlled processes is strongly affected by the physical growth of the medium in which they take place. This observation has far-reaching implications for the kinetics of many encounter-controlled processes, notably in biology and cosmology. Despite the wealth of literature on diffusion-controlled reactions and reaction-diffusion processes in biological systems, most studies accounting for domain growth appear to be focused on pattern formation. One exception are recent models on morphogen gradient formation relying on a diffusion equation similar to Eq.~\eqref{fpe}, possibly augmented with a linear loss term describing morphogen degradation \cite{Averbukh2014} or with additional terms that describe more complex processes affecting the concentration of morphogens \cite{Fried2015}. In these cases, one is led to solve the relevant set of PDEs numerically.

From a more general perspective, reaction-diffusion models for systems of varying size are expected to play an important role when dealing with many problems related to the field of developmental biology, notably those involving growth by cell division \cite{Murray2003}. However, a word of caution is in order here. First of all, in view of the difficulties encountered (even in the one-dimensional case), it is not guaranteed that analytic solutions can be obtained for any boundary value problem of interest; thus, in many situations one will be quickly led to resort to numerical methods. Secondly, the inclusion of additional reaction terms in the relevant PDEs complicates the solution extraordinarily, even if one restricts oneself to linear birth and death processes. Finally, we have seen that PDE approaches fail to capture fluctuation effects, which are in general non-negligible owing to the relevance of noise-induced phenomena in biological systems. Thus, successful approaches to the problem of reaction-diffusion processes in growing domains call for a combination of different techniques allowing one to attain a comprehensive understanding on the basis of multiple, complementary insights.

\section*{Acknowledgments}

\hspace{0.6cm} C.~E. is grateful to the Departamento de F\'{\i}sica of the Universidad de Extremadura for its hospitality.
This work was partially funded by MINECO (Spain) through Grants No. MTM2015-72907-EXP (C.~E.) and
No. FIS2016-76359-P (partially financed by FEDER funds) (S.~B.~Y. and E.~A.), and by the Junta de Extremadura through Grant No. GR15104 (S.~B.~Y. and E.~A.). F. ~L.~V. acknowledges generous financial support from the Fundaci\'on Tatiana P\'erez de Guzm\'an El Bueno and from the Junta de Extremadura through Grant. No. PD16010 (FSE funds).

\bibliographystyle{acm}
\bibliography{AmasAchap}

\end{document}